\documentclass[a4paper,10pt]{article}
\usepackage{subfig,mathtools,url}
\usepackage[affil-it]{authblk}
\pdfoutput=1

\begin{document}

\title{Correlations and Flow of Information between The New York Times and Stock Markets.}

\author[1]{\small Andr{\'e}s Garc{\'i}a-Medina\thanks{Corresponding author~(andgarm.n@gmail.com)}}
\author[2]{\small Leonidas Sandoval Junior}
\author[1]{\small Efra{\'i}n Urrutia Ba{\~n}uelos}
\author[3]{\small A. M. Mart{\'i}nez-Arg{\"u}ello}

\affil[1]{\footnotesize Physics Research Department, Sonora University, Hermosillo, Sonora 83000, M\' exico.}
\affil[2]{\footnotesize Insper, Instituto de Ensino e Pesquisa, Rua Quat\'a, 300, S{\~a}o Paulo, SP, 04546-2400, Brazil.}
\affil[3]{\footnotesize Instituto de F\'isica, Benem\'erita Universidad Aut\'onoma de Puebla, Apartado Postal J-48, Puebla 72570, Mexico}

\maketitle

\begin{abstract}
We use Random Matrix Theory~(RMT) and information theory to analyze the correlations and flow of information between 64,939 news from \emph{The New York Times} and 40 world financial indices during 10 months along the period 2015-2016. 
The set of news was quantified and transformed into daily polarity time series using tools from sentiment analysis.
Results from RMT shows that a common factor lead the world indices and news, and even share the same dynamics. Furthermore, the global correlation structure has found preserved when adding white noise, which indicate that correlations are not due to sample size effects. 
Likewise, we found a lot of information flowing from news to world indices for specific delay, being of practical interest for trading purpose.
Our results suggest a deep relationship between news and world indices, and show a situation where news drive world market movements, giving a new evidence to support  behavioral finance as the current economic paradigm. 
\end{abstract}

\begin{center}
\end{center}

\section{Introduction}

The purpose of this work is to understand, in the context of Econophysics~\cite{Mantenga-Stanley:2000,Bouchaud-Potters:2000,Voit:2005}, the validity of the relatively new school of though named behavioral finance and contrast it with the most accepted paradigm of the efficient market hypothesis~(EMH), on which most of the current financial models rely~\cite{Fama:1998, Shiller:2003, Barberis:2003}.
According to EMH, the stock price instantly incorporates all available market information, and its value does not depend on the price in the past~\cite {Fama:1965}.
However, recently a series of works have begun to investigate the influence of textual sources from Internet to market movements~\cite{Zhang:2010, Bollen:2011, Smailovic:2013, Oliveira:2013, Preis:2013, Alanyali:2013, Zheludev:2014, Gogas:2015}, showing that information extracted from~\emph{Twitter}, \emph{StockTwits}, \emph{Google Trends}, and some financial magazine as~\emph{Financial Times}, give early indications that may help predict changes in stock market.
These new results are building a strong support against the well accepted efficient market paradigm, and supporting the approximation of behavioral economics.
Nevertheless, the above mentioned works have studied the involved time series in individual manner. On the contrary, our intention is to study the global or common properties  of a set of financial indices to know if the information coming from \emph{The New York Times}~(NYT) contains reliable or true information, that is, far from being noise, which problem has been extensively study in our previous work with \emph{Twitter} data~\cite{Garcia:2016}. In addition to this, we are interested now to know under which circumstances the information flows from NYT to financial index prices.

NYT is a newspaper founded and published in New York City,  with over three million subscriptions to their print and digital products in 195 countries all over the world, which makes it one of the most accessible and widely circulated newspapers worldwide~\cite{NYT:2017}.
In order to quantify the news extracted from NYT we used sentiment analysis~\cite{Pang:2008,Jurafsky:2000}, where the dictionary-based approach was followed due to its low computational cost and good precision compared to the learning machine methods~\cite{VADER:2014}. 
Furthermore, the mood polarity was adopted as the sentiment analysis indicator, because this amount can be directly associated with the positive and negative movements of the financial indices~\cite{Bollen:2011}.

On the other hand, understanding the correlation structure between financial markets and discriminate it from noise is of great interest in the context of portfolio optimization~\cite{Markowitz:1959}. A new approach to understanding such correlation comes from Random Matrix Theory~(RMT). Historically, many phenomena of theoretical physics have been successfully solved using  RMT~\cite{Wishart:1928, Wigner:1955,Mehta:1967,Brody_etal:1981,Guhr_etal:1998}, and remarkably a great number of applications to finance have arisen during the last years~\cite{Stanley_etal:1999,Bouchaud_etal:1999,Plerou:2000,Plerou:2002,Potters:2005,Guhr:2010,Maslov:2001,Wang_etal:2011,Sandoval-Franca:2012,Kumar-Deo:2012}.
Although, the study of correlations is useful to determine which assets behave similarly, using only correlation measures we can not establish a causal relationship or influence among financial indices since the action of one variable on another is not necessarily symmetric~\cite{Sandoval-Entropy:2014}. A very useful amount to measure causal phenomenon has foundations in information theory and it is known as transfer entropy. The transfer entropy or transfer information is a dynamic and non-symmetric measure, which was initially developed by Schreiber~\cite{Schreiber:2000}, and is based on the concept of Shannon entropy~\cite{Shannon:1948}. This measure was designed to determine the directionality of transfer information between two processes, by detecting the asymmetry between their interactions~\cite{Schreiber:2000, Prokopenko:2014}. Transfer entropy has been used to solve numerous problems. It has been useful in the study of the neuronal cortex of the brain~\cite{Papana:2011,Shew:2011,Vicente:2011,Faes:2013}, in the study of social networks~\cite{Galstyan:2012}, finance~\cite{Baek:2005,Kwon:2008,Jizba:2012,Sandoval-Entropy:2014}, statistical physics~\cite{Barnett:2012}, and in dynamic systems~\cite{Liang:2013}, receiving a thermodynamic interpretation in~\cite{Prokopenko:2013}.

In this work we use RMT and transfer entropy in order to find out if news drives market movement and to show more evidence against EMH which could support behavioral finance. Specifically, here we analyze the correlations and flow of information between a set of 64,939 news from the New York Times and 40 world financial indices during 10 months along the period 2015-2016.

The paper is organized as follow. Section 2 describes briefly the analyzed data, the methodology to extract the news from NYT, and how polarity time series are constructed using sentiment analysis. Section 3 contains the main results for correlation analysis via RMT. Section 4 shows the flow information results by transfer entropy. Finally, Section 5 presents the conclusions of the work.

\section{Analyzed data}

We consider a set of 64,939 news from NYT and the daily closing values of 40 countries around the world, obtained for the lapsed period of time from July 1 2015 to May 1 2016, which correspond to $T=217$ trading days.
The set of news was extracted in relation with every country listed in Table~\ref{t.1}.
The Bloomberg symbols of the related world financial indices are also listed in table~\ref{t.1}.
The news extraction was made in the coordinated universal time~(UTC), while the time request of the closing prices varies depending of the time zone where the stock markets trade.\\

\begin{table}
\begin{center}
\caption{\footnotesize List of financial data analyzed in this work. First column: country
where the index is traded; second column: Bloomberg ticker of the financial index; third column: keyword to search through the article search API of The NYT.}  
{\begin{tabular}{ l|l|l}
{\bf Country} & {\bf Bloomberg Ticker}  & {\bf NYT Keyword} \\ 
\hline
United States & SPX & United States \\ 
Canada & SPTSX & Canada \\ 
Mexico & MEXBOL  & Mexico\\ 
Colombia & IGBC  & Colombia\\ 
Venezuela & IBVC  & Venezuela\\ 
Chile & IPSA  & Chile\\ 
Argentina & MERVAL & Argentina\\ 
Brazil & IBOV  & Brazil\\ 
Nigeria & NGSEINDX  & Nigeria\\
United Kingdom & UKX & England\\ 
France & CAC & France\\ 
Belgium & BEL20  & Belgium\\
Italy & FTSEMIB  & Italy\\
Switzerland & SMI & Switzerland\\
Netherlands & AEX & Netherlands\\
Denmark & KFX & Denmark\\
Norway & OBX & Norway\\
Sweden & OMX & Sweden\\
Germany & DAX & Germany\\
Poland & WIG & Poland\\ 
Austria & ATX & Austria\\
Greece & ASE & Greece\\
Hungary & BUX & Hungary\\
Ukraine & PFTS & Ukraine\\
Russia & INDEXCM & Russia\\
Turkey & XU100 & Turkey\\
Egypt & CASE & Egypt\\ 
Israel & TA-25 & Israel\\
Arabia & SASEIDX & Arabia\\
Pakistan & KSE100 & Pakistan\\
India & SENSEX & India\\ 
Indonesia & JCI & Indonesia\\ 
Malaysia & FBMKLCI & Malaysia\\ 
Singapore & FSSTI & Singapore\\ 
China & SHCOMP & China\\ 
Hong Kong & HSI & Hong Kong\\ 
Taiwan & TWSE & Taiwan\\ 
South Korea & KOSPI & South Korea\\  
Japan & NKY & Japan\\ 
Australia & AS51 & Australia
\end{tabular}\label{t.1}}
\end{center}
\end{table}

The news were extracted through the interface~\emph{Article Search API} of NYT, which gives us access to its database in a structured way. In order to remove the noise of the extracted text due to the non-alphanumeric characters, we preprocess it with the help of the Natural Language Toolkit~(NLTK) of PYTHON. \\

Once removed the noise, we apply the sentiment analysis to the cleaned text using the Valence Aware Dictionary and sEntiment Reasoner (VADER), which is a lexicon that implements syntactical and grammatical rules, incorporating empirically derived quantifiers in order to take into account the sentiment intensity present in the analyzed text, where every element of the lexicon is scoring between -4 and 4, from very negative word to very positive word. VADER has been adjusted to capture the sentiment expressed in social networks, but also has been shown excellent results capturing the sentiment of NYT text~\cite{VADER:2014}.\\

Finally, we took the average of all the scores found in a given text~(one score for each word), and we link this average score as the emotional polarity~(positive or negative) of it~\cite{Garcia:2016,Bollen:2011}. Then, we named as $P_k(t)$ the polarity time series of all the news corresponding to a given keyword $k$ at a given day $t$.

\section{Random matrix theory analysis}
\subsection{Preliminaries}
\label{RMT}

Let $S_k(t)$ denote the daily closing prices of index $k$ at day $t$.
The return value for each index $k$~($k=1,\dots,N$) at times $t=1,\dots,T$~(measure in days) is obtained by
\begin{equation}
\label{eq.2}
R_k(t) = \frac{S_k(t+\Delta t) -S_k(t)}{S_k(t)}.
\end{equation}
We choose a return interval of one trading day, such that $\Delta t = 1$. 
In order to compare our data with the universal results of RMT, both polarity and return time series are normalized.
The respective normalized return, for index $k$ at time $t$, is defined as 
\begin{equation}
\label{eq.3}
r_k (t)  = (R_k (t) - \langle R_k \rangle )/\sigma_k,
\end{equation}
\noindent
where $\sigma_k$ is the standard deviation of $R_k$, and $\langle \dots \rangle$ denotes the time average over the studied period.
The average polarity is normalized in the same way and is denoted as $p_k (t)$ for the index $k$ at time $t$.\\

\begin{figure}
\centering{\includegraphics[width=0.9\linewidth]{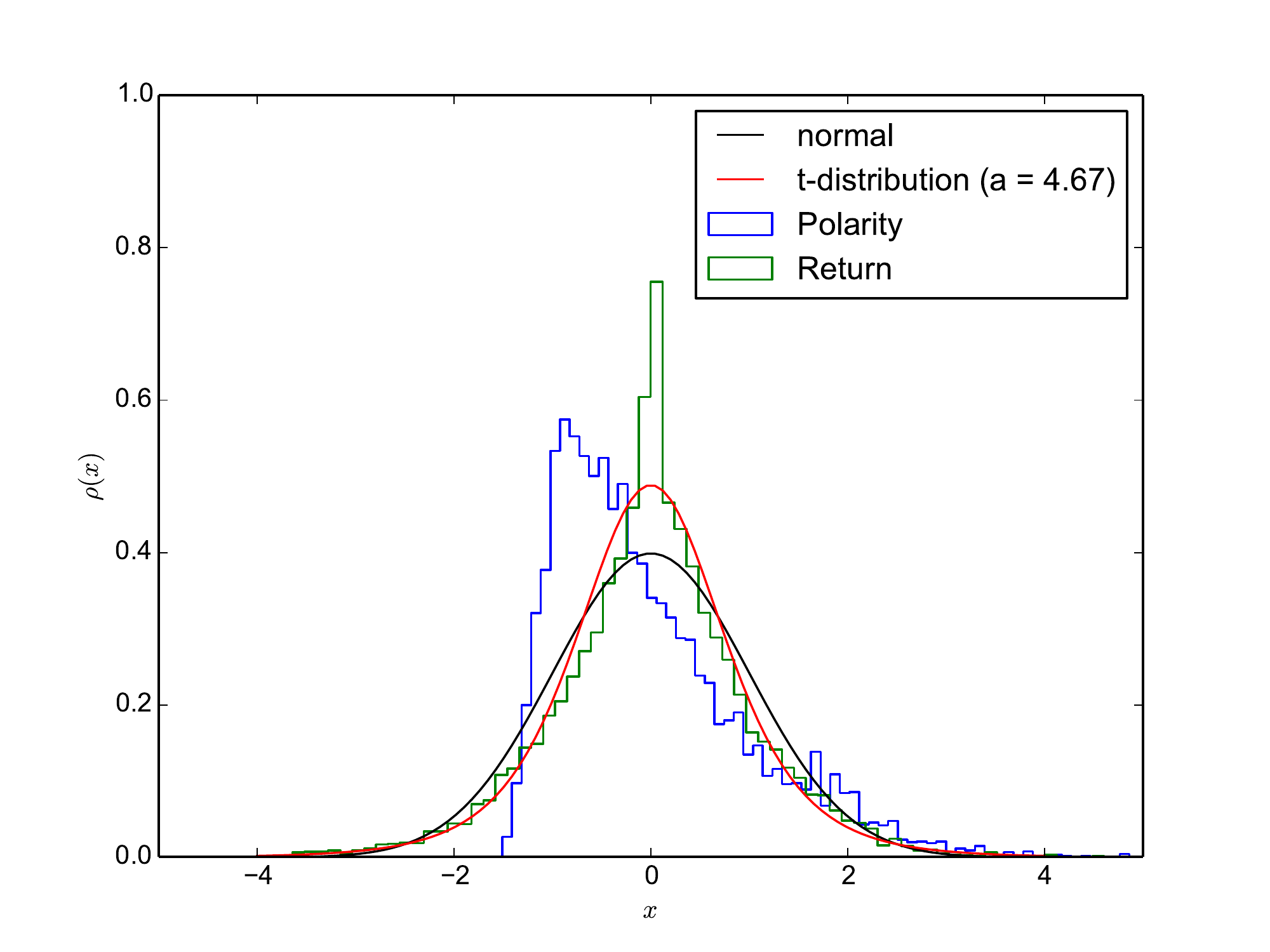}}
\caption{\footnotesize Distribution of returns, polarities, as well as the normalized Gaussian distribution and the student-t distribution with the parameter $a=4.67$, which best fits the return distribution.}
\label{distribution}
\end{figure}

In Fig.~\ref{distribution} we plot the distribution of returns, polarities, as well as the normal distribution and the student-t distribution with parameter $a=4.67$, which best fit the return distribution. It is know that the distribution of returns  usually has tails heavier than the tails of the normal distribution~\cite{Mantenga-Stanley:2000} and is better characterizing by the student-t distribution, behavior that we can observe in  the same Fig.~\ref{distribution}. On the contrary, it can be seen that the polarity distribution has an skewed shape, which rule out the symmetric behavior found for return data.\\

\subsection{Wishart Ensemble}

The correlation matrix element between demeaned and standardized time series $x_k$ and $x_l$ is given by
\begin{equation}
\label{eq.4}
c^{(x)}_{k,l} = \langle x_k(t) x_l(t) \rangle,
\end{equation}
where $x$ denotes the type of time series which we are working on, such that $c^{(p)}_{k,l}$ and $c^{(r)}_{k,l}$ are the matrix correlation elements constructed from polarity and return time series, respectively.\\ 

Let be $W$ an $N\times T$ matrix whose matrix elements are statistically independent Gaussian variables with zero mean and equal variance. 
Then, the matrix $H = WW^\dagger$  is known into the formalism of RMT as Wishart matrix and a set of these matrices under the Haar measure as Wishart ensemble (WE)~\cite{Wishart:1928}. 
By construction, these matrices are formed with $N$ uncorrelated time series of finite length $T$. 
The $N$ eigenvalues of $H$, denoted by ${\lambda_1,\lambda_2,\dots,\lambda_N}$, are non-negative and have the joint probability density function given by~\cite{Majumdar:2011}
\begin{equation}
P[\{\lambda_i\}] = C_{N,T} \exp \left[ -\frac{\beta}{2} \sum_i^N \lambda_i \right]
\prod_{i = 1}^N \lambda_i^{\alpha\beta/2} \prod_{j<k} |\lambda_j - \lambda_k|^{\beta},
\label{Jam641}
\end{equation}
where $\alpha = (1 + T - N) - 2\beta$, and the normalization constant $C_{N,T}$ can be computed
exactly~\cite{James:1964}. We can assume $N \leq T$, because if $N \geq T$, one can show that $N-T$ eigenvalues are exactly
0 and the rest of the $T$ eigenvalues are distributed exactly as in the above expression with $N$ and $T$ exchanged.
The solution for $\beta=1$~(real symmetric case) in the limit $N, T \rightarrow \infty$, with $Q = T/N (\geq 1)$, is given by the Mar\v{c}enko-Pastur law~\cite{M-P:1967}
\begin{equation}
\label{eq.7}
\rho(\lambda) = \frac{Q}{2\pi\sigma^2} \frac{\sqrt{(\lambda_{+} - \lambda)(\lambda - \lambda_{-})}}{\lambda},
\end{equation}
\noindent
within the bounds $\lambda_{-} \leq \lambda \leq \lambda_{+}$ and $0$ otherwise. The smallest~(largest) eigenvalue of a random matrix in WE is given by~\cite{M-P:1967}:
\begin{equation}
\label{eq.8}		
\lambda^{+}_{-} = \sigma^2(1+1/Q\pm 2\sqrt{1/Q}).
\end{equation}
These predictions are known in RMT field as universal results of Wishart matrices, and make up the null hypothesis of no correlations between financial indices (or polarities). If there is no correlation between financial indices (or polarities), then the eigenvalues should be bounded between RMT predictions\cite{Stanley_etal:1999,Bouchaud_etal:1999}.\\

\begin{figure}
\centering
\subfloat{\includegraphics[width=0.9\linewidth]{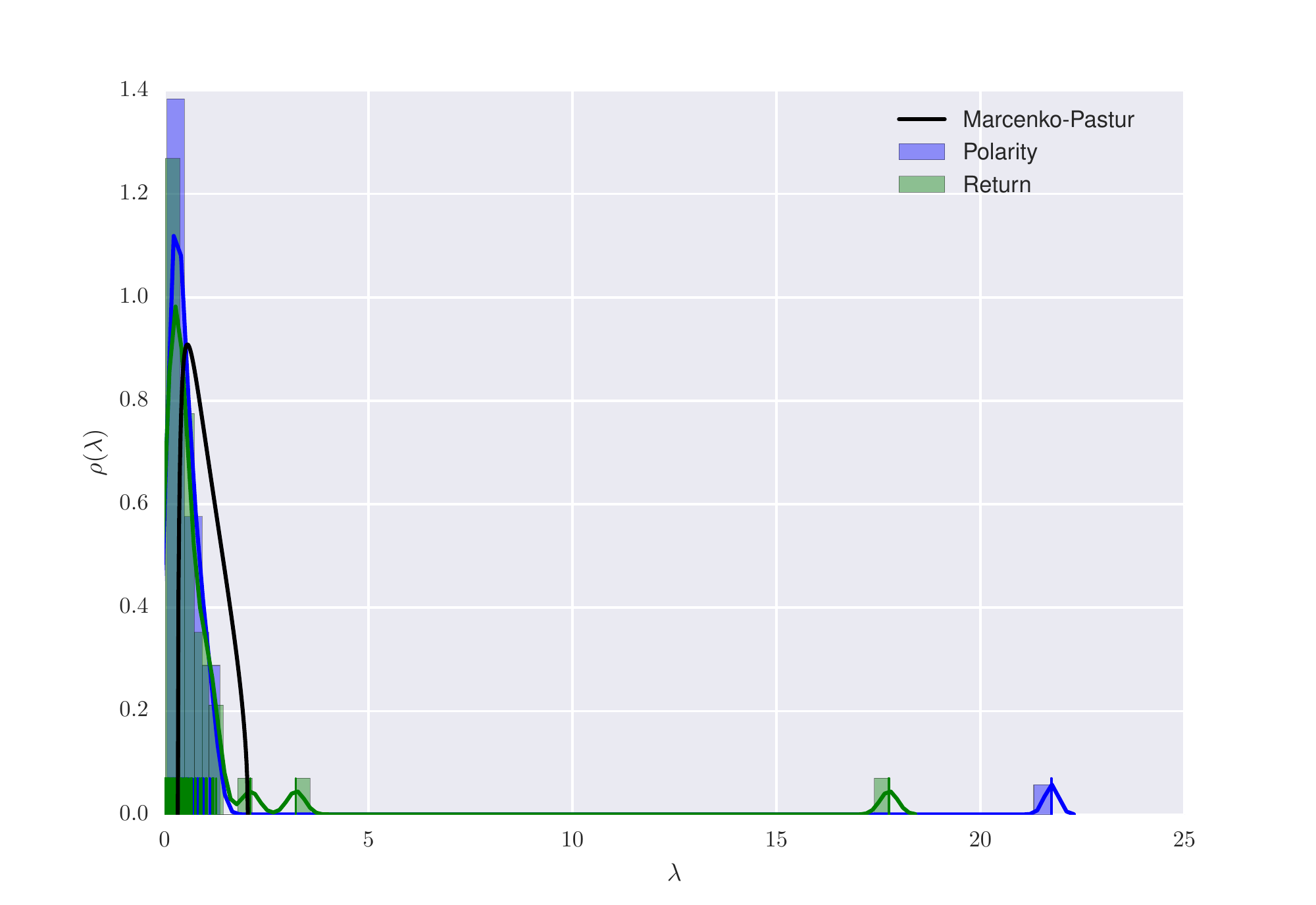}}
\caption{\footnotesize Eigenvalue distribution of correlation matrices. We show with black line the Mar\v{c}enko-Pastur law. The blue line represent the results for polarities and the green line for returns.}
\label{fig.3}
\end{figure}

In Fig.~\ref{fig.3} we plotted the eigenvalue distribution for the empirical correlation matrices $C^{(r)}$ and $C^{(p)}$, with the Mar\v{c}enko-Pastur law superimposed on them. Here the dimensions are $N\times T=40 \times 217$ with parameter $Q=T/N=5.425$.
We clearly see some eigenvalues far away the upper bound of the noise zone for both correlation matrices, which give information about the correlation behavior of the whole set of countries under study, and implies the presence of true correlation. We will discuss more about the largest eigenvalue and the corresponding largest eigenvector in the next section.\\ 

\subsection{Eigenvectors and Temporal Analysis}

In finance, the fact that all the components of the eigenvector associated to the largest eigenvalue are positive reflects a common financial market mode and it is related with the most risky mix of assets in a portfolio of investment. On the contrary, the eigenvector associated to the smallest eigenvalue correspond to the less risky portfolio. These two eigenvectors are not random combinations of variables~\cite{Bouchaud:2011}.\\

\begin{figure}
\centering
\subfloat[]{\includegraphics[width=0.9\linewidth]{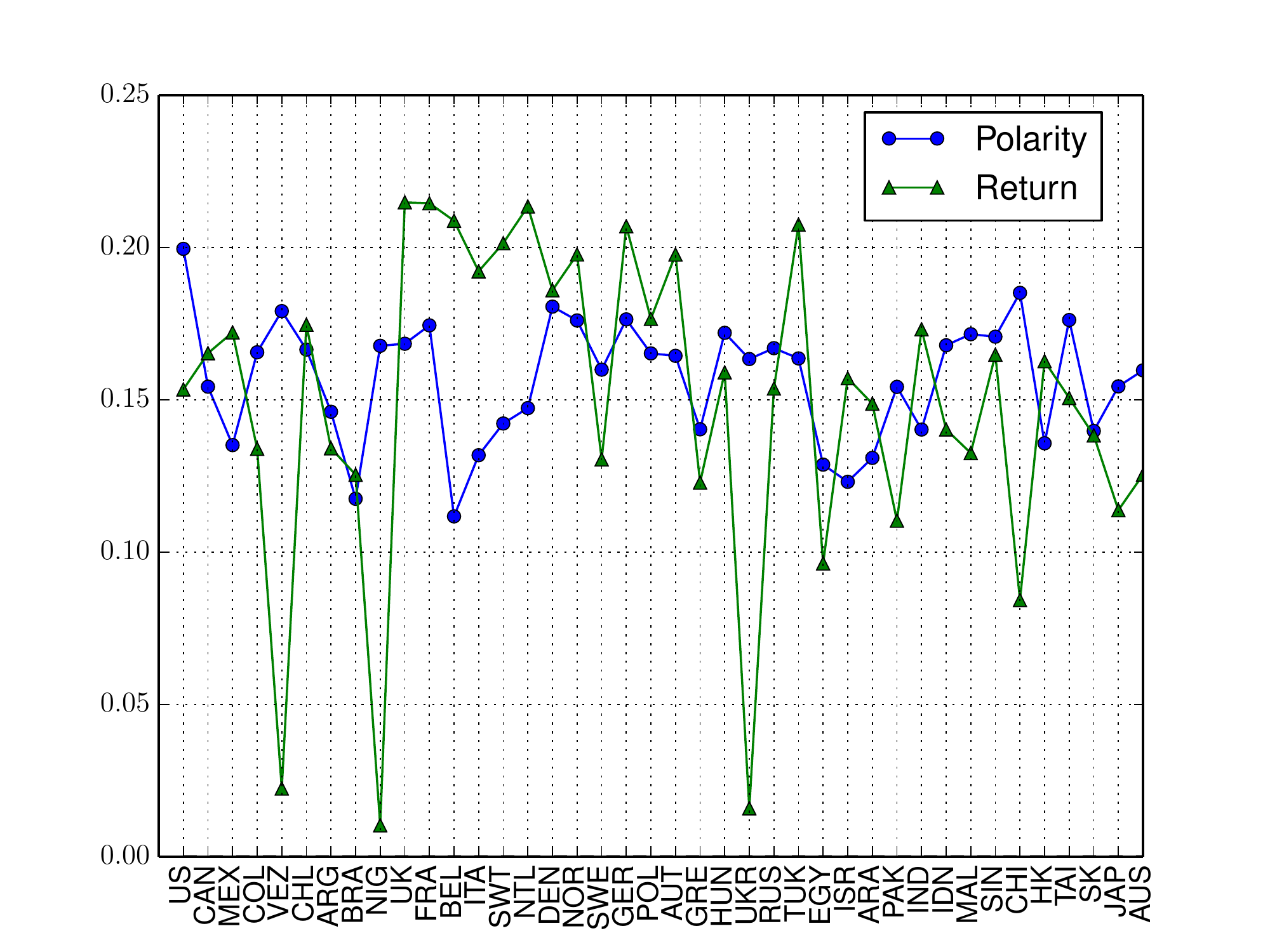}}\\
\subfloat[]{\includegraphics[width=0.9\linewidth]{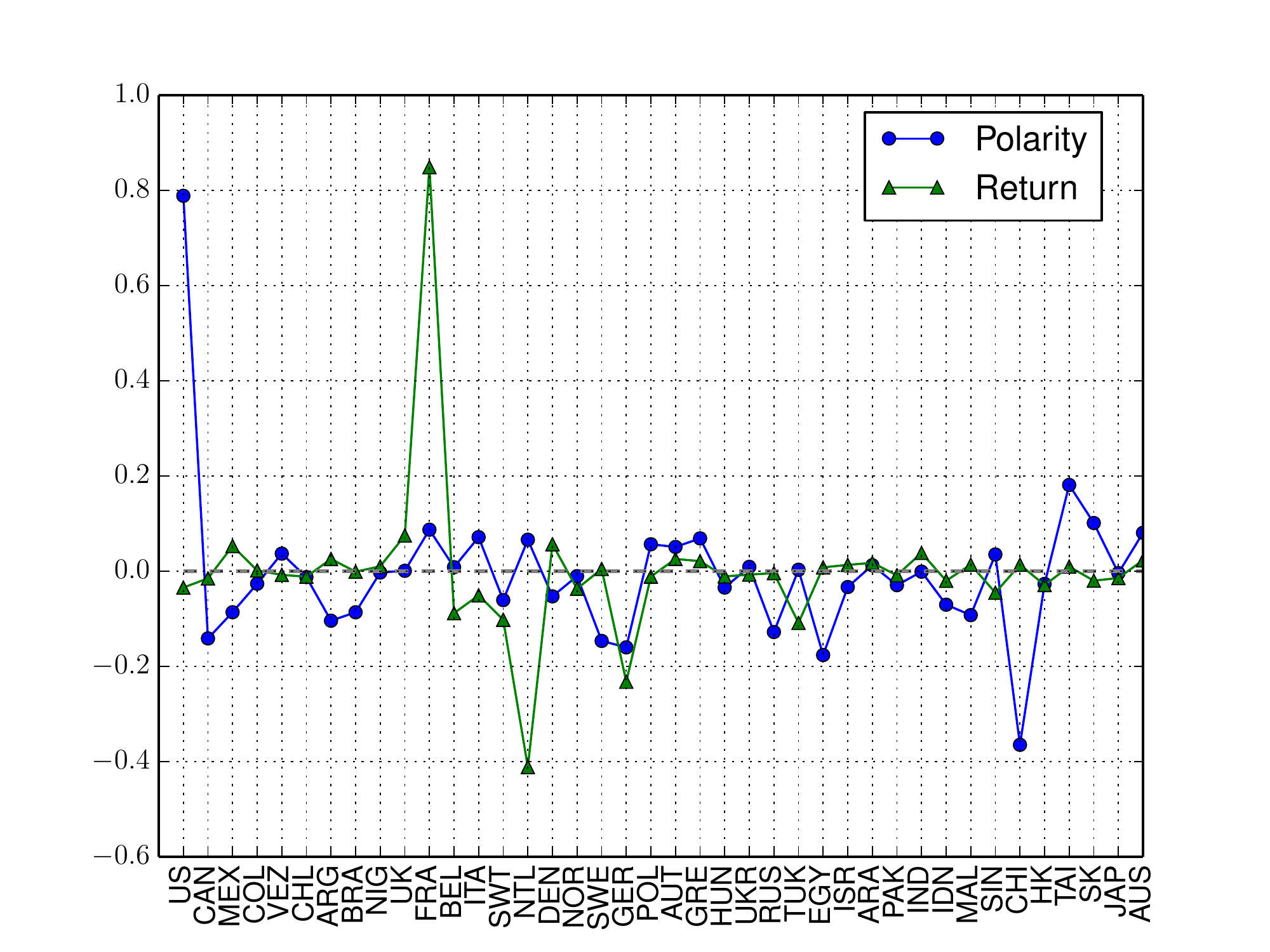}}
\caption{\footnotesize Eigenvector components of the associated largest (a) and smallest (b) eigenvalue of the empirical correlation matrices.}
\label{eigenvector}
\end{figure}

In Fig.~(\ref{eigenvector}) we have plotted the eigenvectors associated to the largest and smallest eigenvalue of the empirical correlation matrices. We can see in Fig.~\ref{eigenvector}(a) that all the components of the eigenvector associated to the larger eigenvalue are positive and far from zero, while in Fig.~\ref{eigenvector}(b) most of the components of the eigenvector associated to the smallest eigenvalue are near to zero and only few elements stand out from them. Both cases behave as expected~\cite{Bouchaud:2011}. Interestingly, if we analyze Fig.~\ref{eigenvector}(b) more closely  it  tell us that the less risky portfolio  must include United States and China if we consider the polarity data to its construction, and also must include France and Netherlands if we consider the return data. We argue this because the magnitude of these components are bigger than the others and then are the most representative of the eigenvector associated to the smallest eigenvalue. \\

To make a temporal analysis, we constructed a set of sample correlation matrices from a sliding window of four trading months ($T_s=$ 160 days) with an overlap of one trading day.
Then, we obtained two sets of $M=58$ sample correlation matrices, one set from polarity values and other set from return values.
Every correlation matrix within these sets has now a dimension $N \times T_s = 40 \times 160$ and parameter $Q=T_s/N=4$.
Consequently, the upper and lower theoretical bounds are $\lambda_{-} = 0.25$ and $\lambda_{+} = 2.25$, for the bulk of the eigenvalue distribution due to noise.\\

\begin{figure}
\centering
\subfloat{\includegraphics[width=0.9\linewidth]{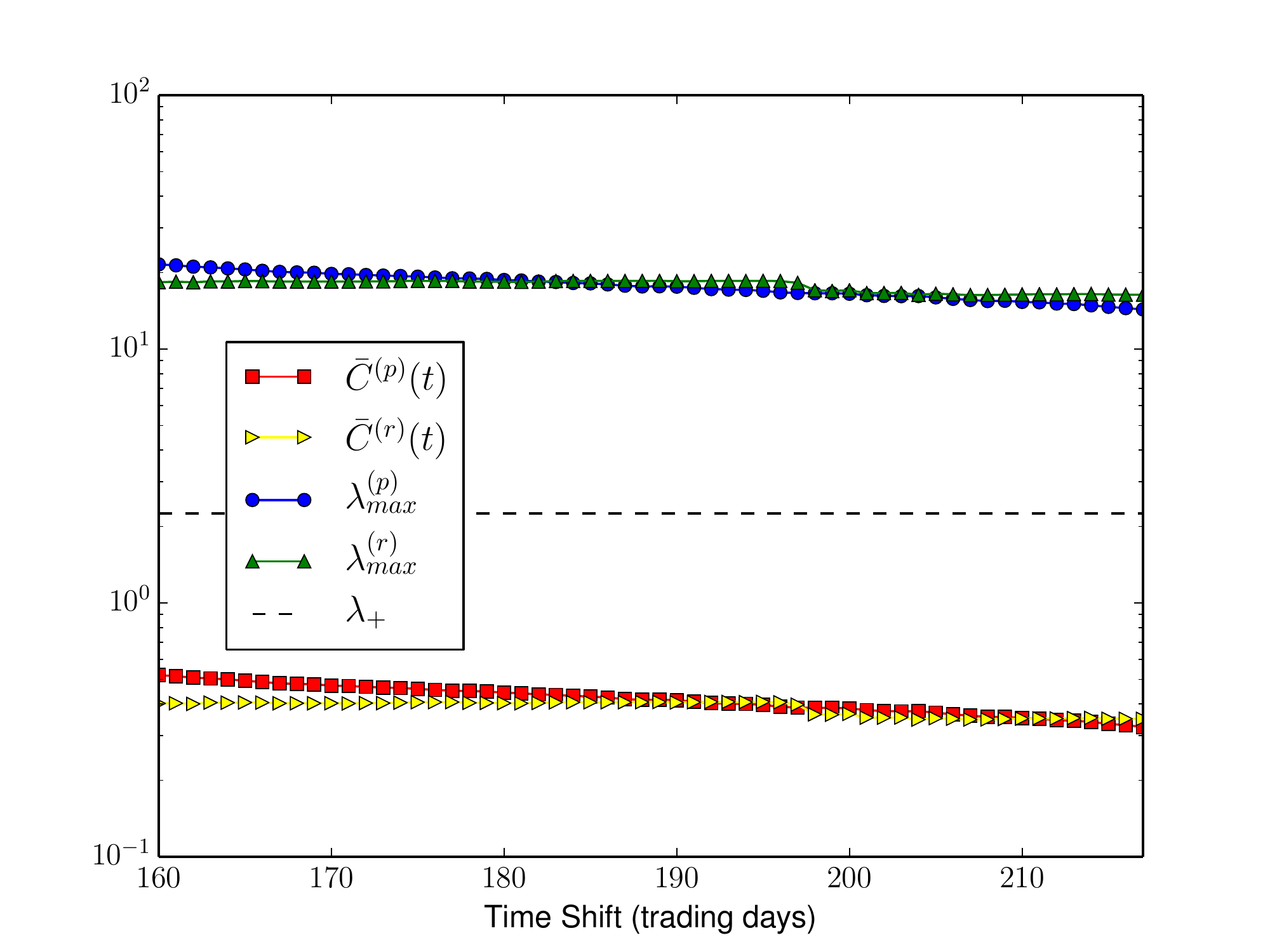}}
\caption{\footnotesize Temporal behavior of $\lambda_{max}(t)$ and  $\bar{c}(t)$ for both empirical data. The dashed line represents the upper theoretical bound of the eigenvalue distribution.}
\label{timevolution}
\end{figure}

It is know that for $C$ of  large dimensions the time evolution of the largest eigenvalue $\lambda_{max}(t)$ is strongly correlated with the mean correlation coefficient $\bar{c}(t) = \langle C(t) \rangle_{ij}$~\cite{Mantegna:2011,Stepanov:2015}. 
In Fig.~\ref{timevolution} we plotted these quantities for our empirical data.
Although the dimension of the empirical data is too small, it has been found a strong correlation between $\lambda_{max}(t)$ and  $\bar{c}(t)$ with Pearson $Pc$ and Spearman $Sc$ correlation coefficients~\cite{Pearson:1895,Spearman:1904} bigger than 0.97 for both polarity and return values.
Then, $\lambda_{max}(t)$ and  $\bar{c}(t)$ share the same dynamics.
Furthermore,  $\lambda^{p}_{max}(t)$  and $\lambda^{r}_{max}(t)$ present a strong correlation of $Pc=0.80$ and $Sc=0.57$.
Then, we can argue that the set of news from NYT and the corresponding global returns also share the same dynamics, supporting the behavioral finance assumptions.\\

On the other side, a simple way to extract information from eigenvectors is by computing the Inverse Participation Ratio (IPR), which allows us to know the number of components that participate significantly in each eigenvector (or portfolio). It exhibits a distinction between the eigenvectors associated to the extreme eigenvalues and the ones associated to the bulk in the noise zone.
The IPR of eigenvector $V^k$ is given by~\cite{Clark:2015}
\begin{equation}
\label{eq.9}
IPR_k = \sum_{j=1}^N |{V_j^k}|^4.
\end{equation}
This quantity always falls between the limits $1/N$ and one.  
It is expected that the values for $IPR_N$ fluctuate near the lower limit $1/N$  because it corresponds to the most diversified portfolio, whereas for $IPR_1$ it is expected higher values because it correspond to the smallest eigenvalue and therefore to the less diversified portfolio~\cite{Markowitz:1959}. Furthermore, for values of $k$ within the region considered as noise it is expected random combinations of assets and then values of $IPR_k$ between  $IPR_N$ and $IPR_1$. \\

\begin{figure}
\centering
\subfloat[]{\includegraphics[width=0.9\linewidth]{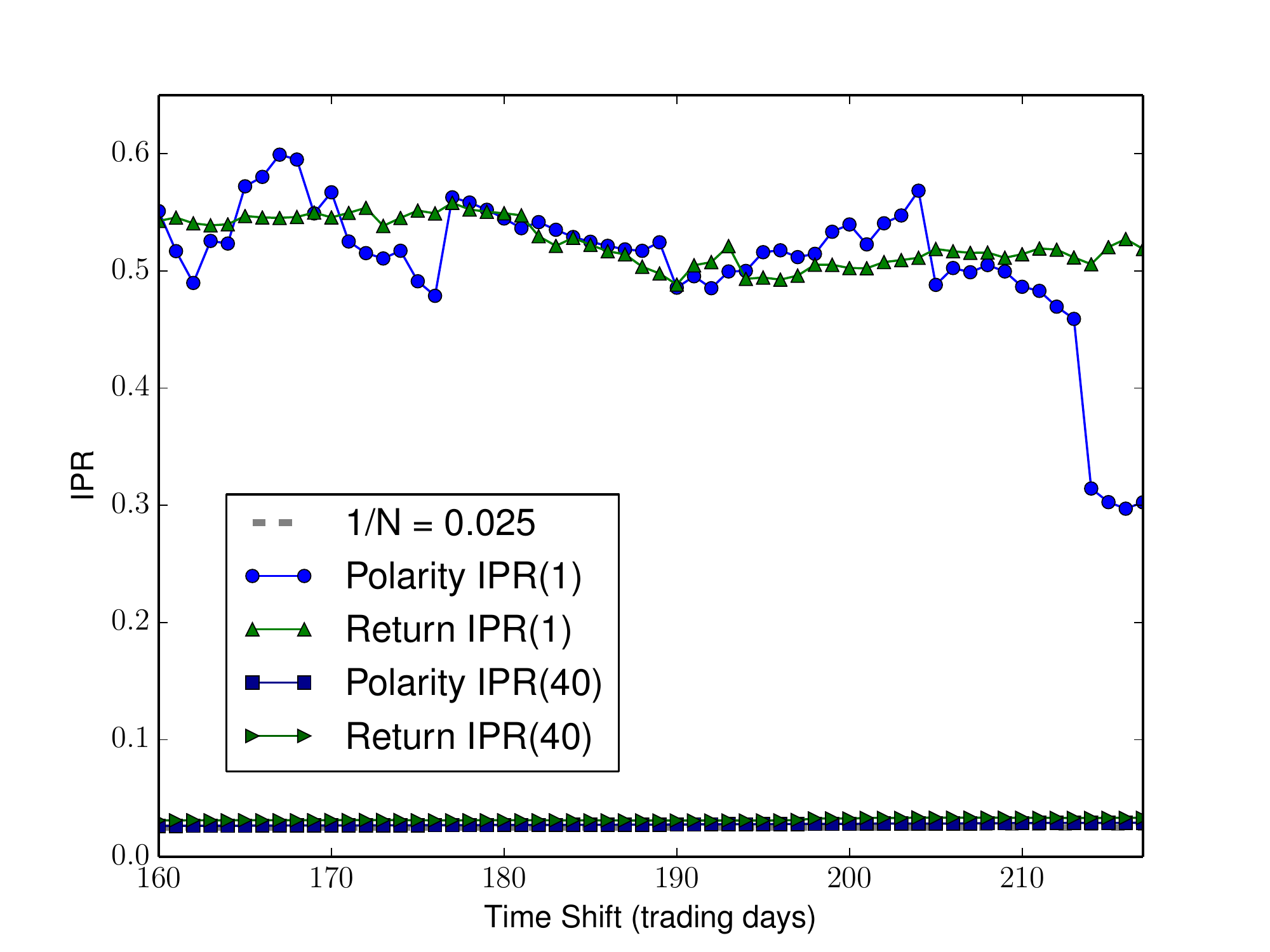}}
\caption{\footnotesize Temporal behavior of the inverse participation ratio corresponding to the largest~(IPR(40)) and smallest~(IPR(1)) eigenvalue. The dashed line represent the theoretical lower bound.}
\label{Inverse}
\end{figure}

In Fig.~\ref{Inverse} are plotted the temporal behavior of $IPR_1$ and $IPR_{40}$.
We obtained a good correlation coefficient between the empirical data in $IPR_{40}$, with $Pc=0.84$ and $Sc=0.54$. The other case $IPR_1$ presents low correlation coefficient of $Pc=0.27$ and $Sc=0.36$ between the temporal behavior of the corresponding eigenvectors. The first results for $IPR_{40}$ confirms the fact that each financial index participates significantly in the eigenvector $V_{40}$, and consequently all the indices move as a whole in this eigenmode. Surprisingly, the data from NYT shows the same behavior. 

\subsection{Correlated Wishart Ensemble}

We are now interesting to know, firstly, if the correlation structure of the empirical data is preserved when adding with noise. Secondly, we desire to characterize the cross-correlations between polarities and returns. A theoretical technique from RMT to analyze these problems is the non-symmetric correlation matrix approach for the Correlated Wishart Orthogonal Ensemble (CWOE)~\cite{Vinayak:2013}.\\ 

Lets start defining CWOE as an ensemble of real symmetric matrices of type $C = \mathcal{W}\mathcal{W}^{t}/T$ , where $\mathcal{W} = \xi^{1/2}W$, $\xi$ is a positive definite nonrandom matrix, and entries of $W$ are independent Gaussian variables with zero mean and variance equal one, i.e.,white noise. Be $D^{(r)}$ and $D^{(p)}$ the data matrices composed of return and polarity time series, respectively. Then we can construct a partitioned data matrix of dimensions $2N$x$T$
\begin{equation}
D = \left(
\begin{array}{c}
D^{(r)}\\
D^{(p)}\\
\end{array}
\right), 
\end{equation}
and a partitioned correlation matrix $C$ defined in terms of 4 blocks
\begin{equation}
C = \frac{1}{T} D D^{T} = 
\left(
  \begin{array}{cccc}
    C^{(r)}   &  C^{(r,p)}  \\ 
    C^{(p,r)}    &   C^{(p)} \\
  \end{array}\right).
  \label{supercorr}
\end{equation}
The diagonal blocks account for the return and polarity correlations alone, further the off-diagonal blocks account for the mixture correlations between returns and polarities, satisfying the relation $C^{(r,p)} = C^{(p,r)T} = D^{(r)}D^{(p)T}$.\\

Likewise, let set $\xi$ as our empirical correlation matrices $C$'s. In this way, $\xi^{(r)} = C^{(r)}$, $\xi^{(p)} = C^{(p)}$, and be $W_{1},W_{2} \in \mathbf{R}^{N\times T}$ two independent Gaussian variables with zero mean and variance one. We can define the partitioned data matrix with white noise $\mathcal{W}$ of dimensions $2N\times T$, constituted by both return and polarity data sets, as
\begin{equation}
\mathcal{W} = \left(
\begin{array}{c}
\sqrt{C^{(r)}} W_{1}\\
\sqrt{C^{(p)}} W_{2}\\
\end{array}
\right).
\end{equation}
Then, the partitioned correlation matrix with noise is given by

\begin{equation}
C' = \frac{1}{T} \mathcal{W} \mathcal{W}^{\dagger} = 
\left(
    \begin{array}{cccc}
    C^{(r)}  W_1 W_1^t   &  C^{(r,p)} W_{1}W_{2}^t   \\ 
    C^{(p,r)} W_{2} W_{1}^t   &   C^{(p)}  W_2 W_2^t \\
  \end{array}\right).
\end{equation}

\begin{figure}
\centering
\subfloat[C]{\includegraphics[width=0.5\linewidth]{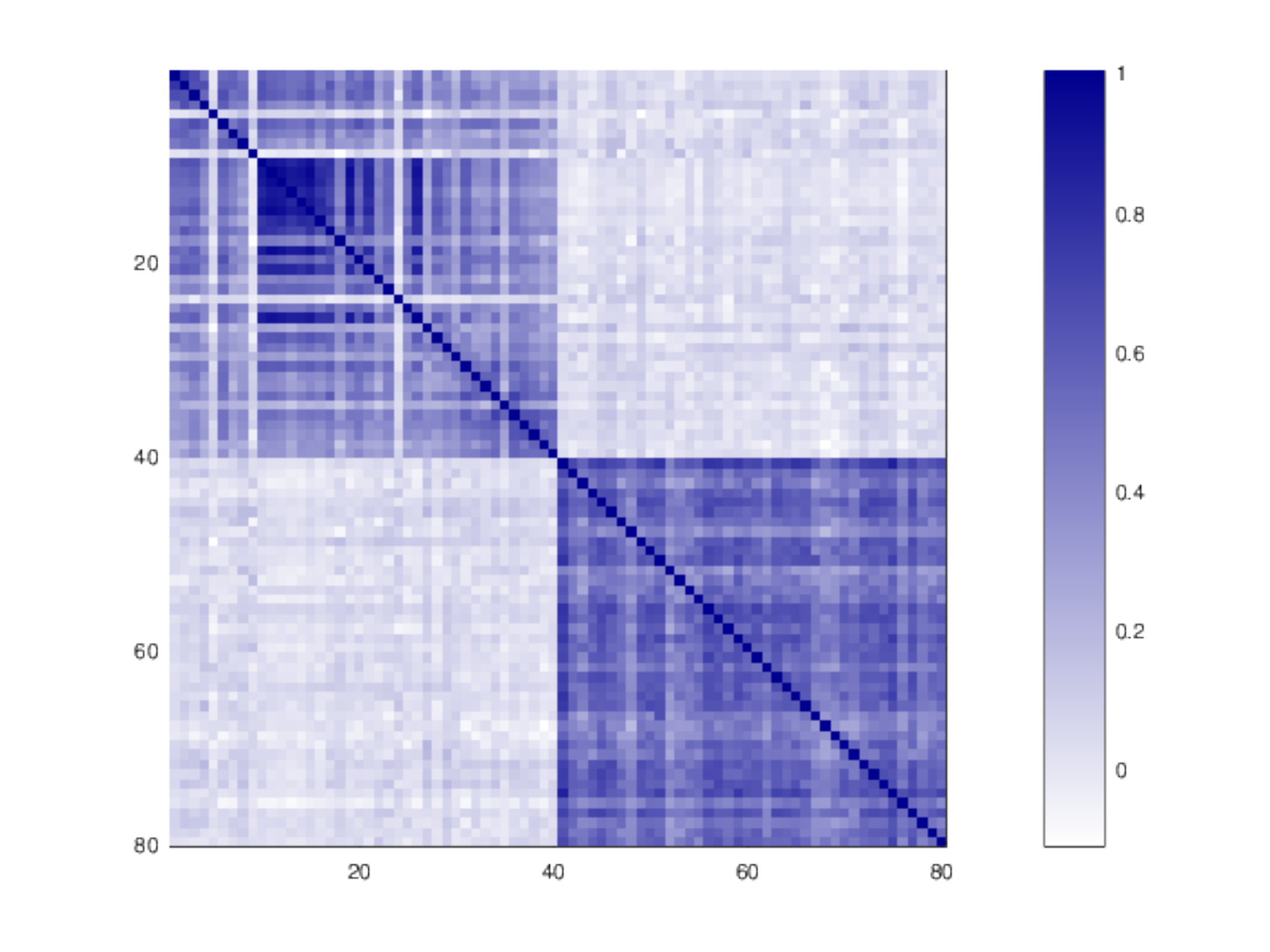}}
\subfloat[C']{\includegraphics[width=0.5\linewidth]{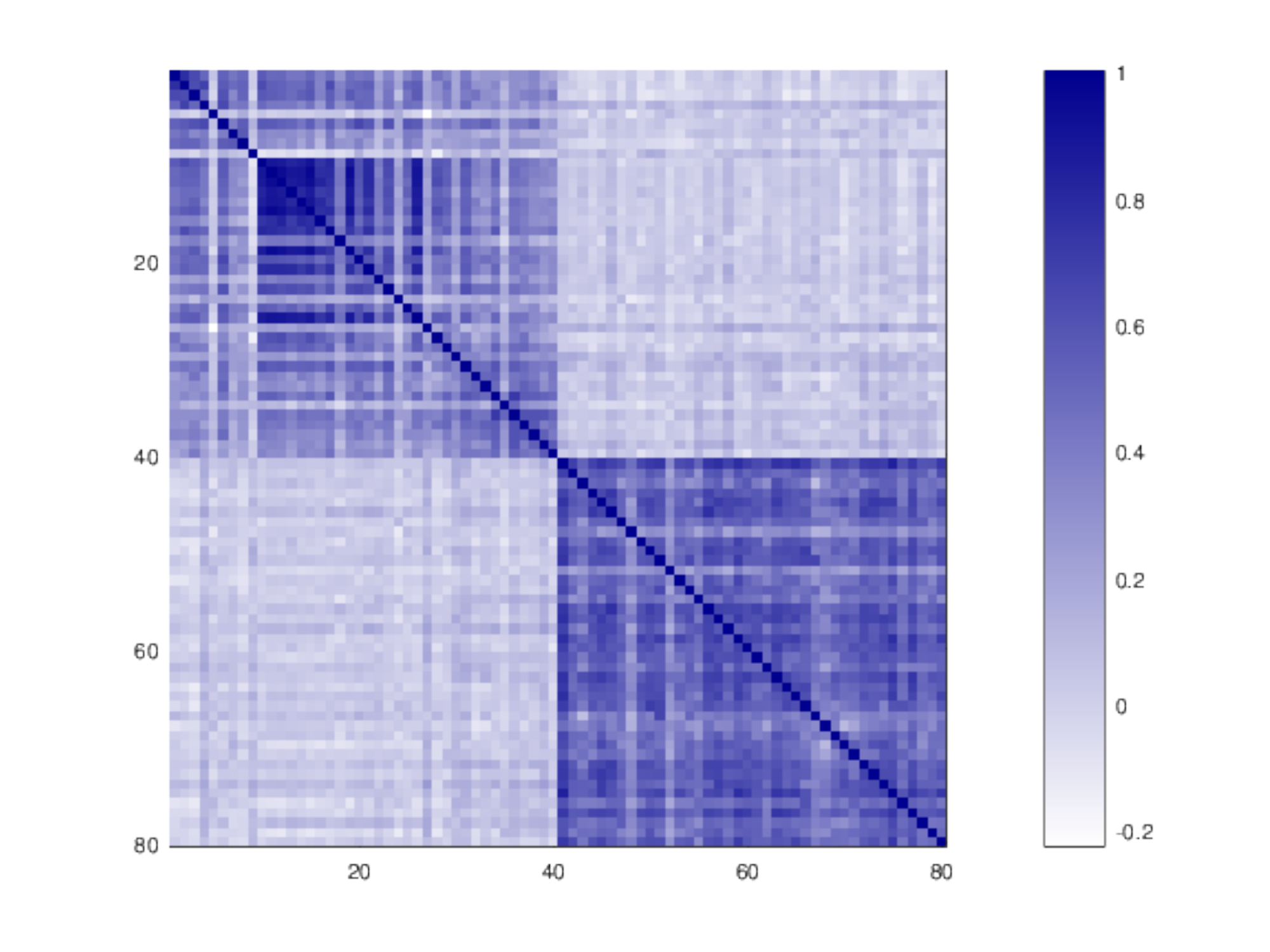}}
\caption{\footnotesize Correlation matrices $C$ and $C'$}	
\label{super}
\end{figure}

We are now in position to inquire about our first goal by comparing the correlation matrices $C$ and $C'$ and measure if the addition of white noise break the correlation structure~(see Fig.~\ref{super}). 
To compare $C$ and $C'$ quantitatively, we calculate the absolute difference of two neighboring correlation coefficients as in~\cite{Schafer:2010}. We found that the average of this value varies less than 10\% if the noise is included.
Then we can argue that the correlation structure of the empirical data carries out more information than white noise and then the correlation structure is preserved as a whole. Therefore, these results support the existence of true correlations in financial indices and polarities.\\

To our second goal, i.e., to characterize the cross-correlations between polarities and returns, we need to look at the off-diagonal blocks of eq.~\ref{supercorr}, which in general are non-symmetric correlation matrices. This kind of matrices have complex eigenvalues. We study the eigenvalues distribution of this matrices in the complex plane by the techniques developed for large dimensions~($N \rightarrow \infty$, $T \rightarrow \infty$) in~\cite{Vinayak:2014}, but the results~(not showed) do not assert the presence of cross-correlations between NYT and the world indices, which might be due to the small dimension~($N=40$) of the empirical correlation matrix. 
To go further we will proceed to study a causality measure instead of linear correlations. \\

\section{Transfer entropy analysis}
\label{TE}

We are now interested in measure the flow of information from news to prices and vice versa via the information theory approach, particularly by the concept of transfer entropy~(TE).
The TE from the variable $Y$ to variable $X$ is given by the expression~\cite{Schreiber:2000}

\begin{equation}
T_{Y\rightarrow X} =\sum_{i_{n + 1}, i_n,j_n = 1}^{n = T-1} p(i_{n+1}, i_n^{(k)}, j_n^{(l)} ) \log \frac{p(i_{n+1} | i_n^{(k)}, j_n^{(l)})}{p(i_{n+1} | i_n^{(k)})},
\label{Transfer}
\end{equation}on 
where $T$ is the time series length of $X$ and $Y$, $P(i,j)$ is the joint probability distribution of both $X$ and $Y$, and $P(i|j)$ is the conditional probability of the variable $X$ given $Y$. 
The last eq.~(\ref{Transfer}) tells us that the element $i_{n + 1}$ of the time series $X$ is influenced by the previous $k$ states of the same time series $X$ and the $l$ previous states of the time series $Y$.\\

We compute $TE$ using the library JIDT~\footnote{Available at http://jlizier.github.io/jidt/}~\cite{JIDT}, which enables to construct the probability distribution functions via a kernel density estimator, defined as~\cite{Wibral-Vicente-Lizier:2014}
\begin{equation}
 p_h = \frac{1}{n} \sum_{i=1}^n K_h (t - t_i),
\end{equation}
where every kernel $K_h$ is identified by the position parameter $t_i$ and the bandwidth $h$.
In our case, the kernel function counts the number of return or polarity values falling inside a box of length $h$ centered at $t_i$.
A very common selection for the $h$ parameter is given by the Silverman's rule~\cite{Silverman:1998}
\begin{equation}
 h = \left(\frac{4\sigma^5}{3n}\right)^{\frac{1}{5}},
 \label{dumb}
\end{equation}
where $\sigma$ is the standard deviation of the time series, and $n$ its length dimension.\\

\begin{figure}
\subfloat[]{\includegraphics[width=0.5\linewidth]{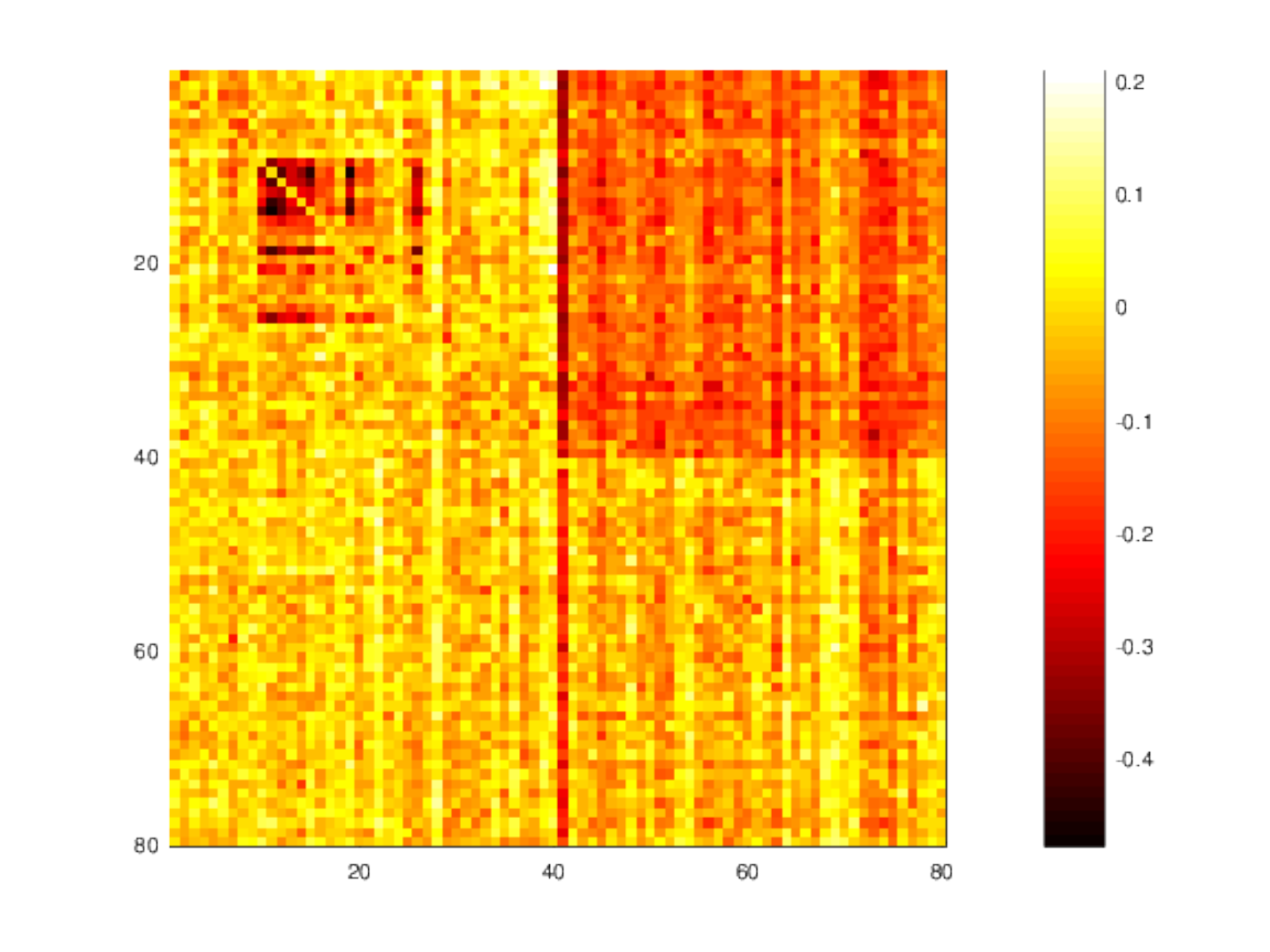}}
\subfloat[]{\includegraphics[width=0.5\linewidth]{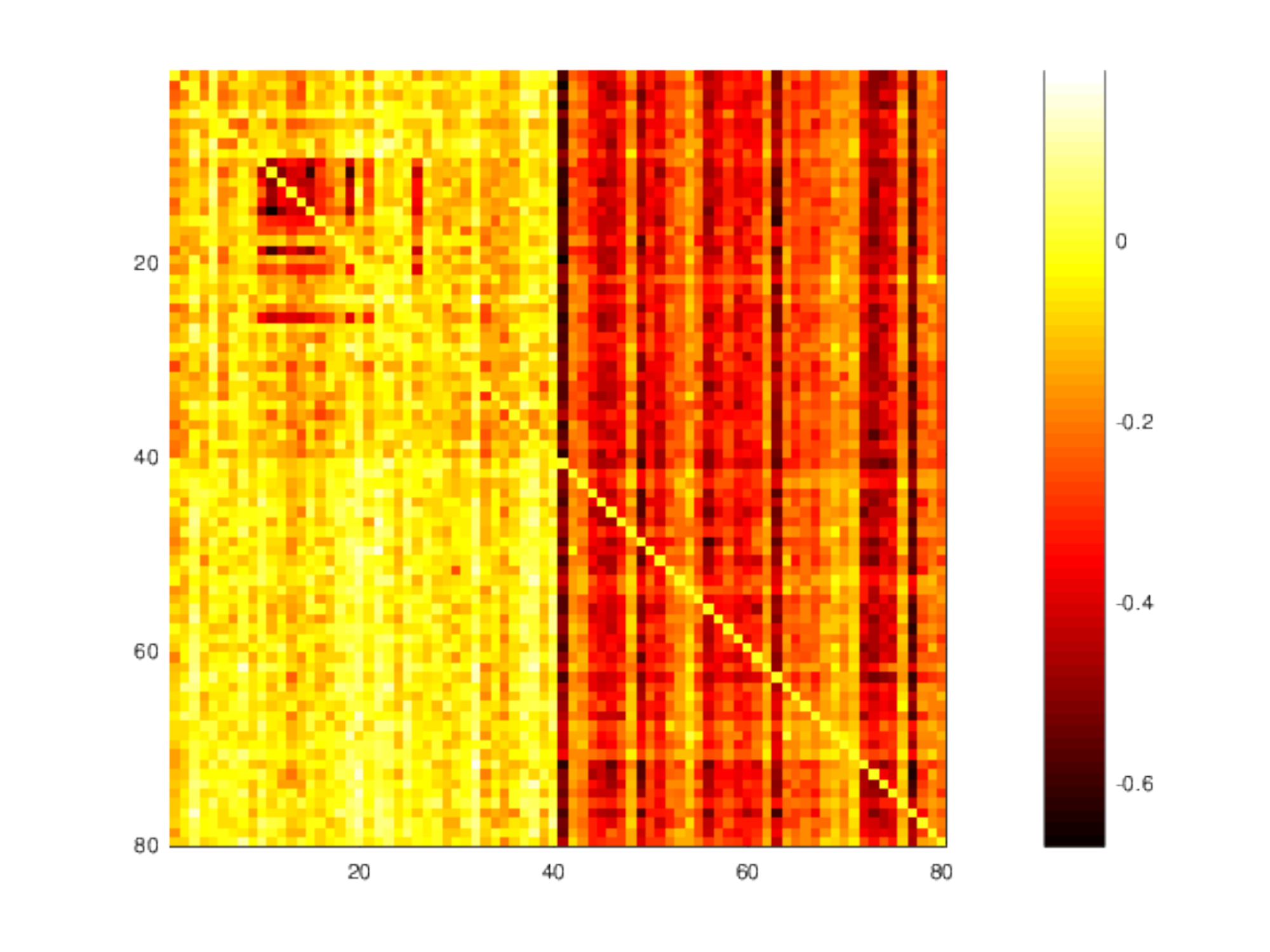}}\\
\subfloat[]{\includegraphics[width=0.5\linewidth]{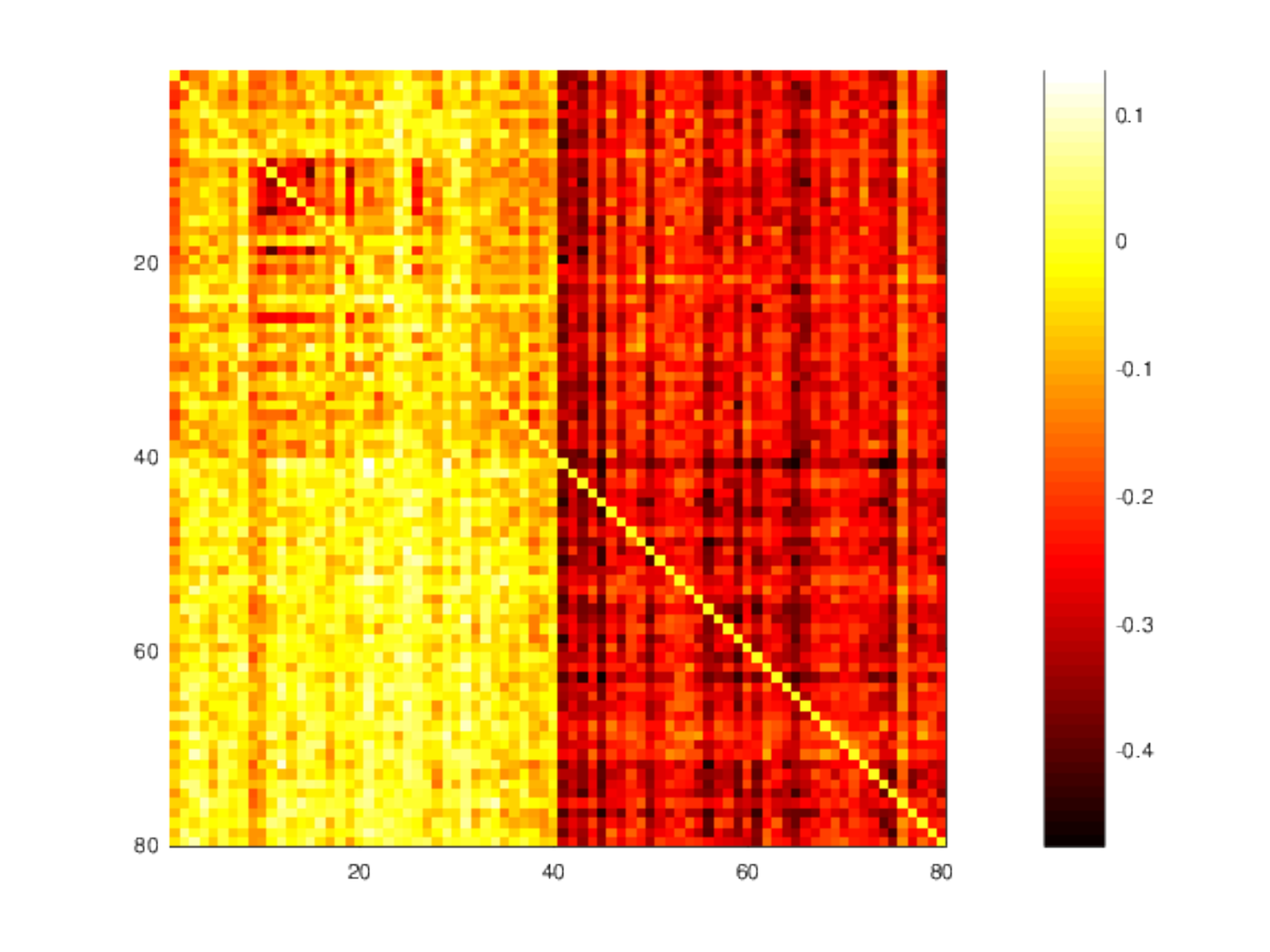}}
\subfloat[]{\includegraphics[width=0.5\linewidth]{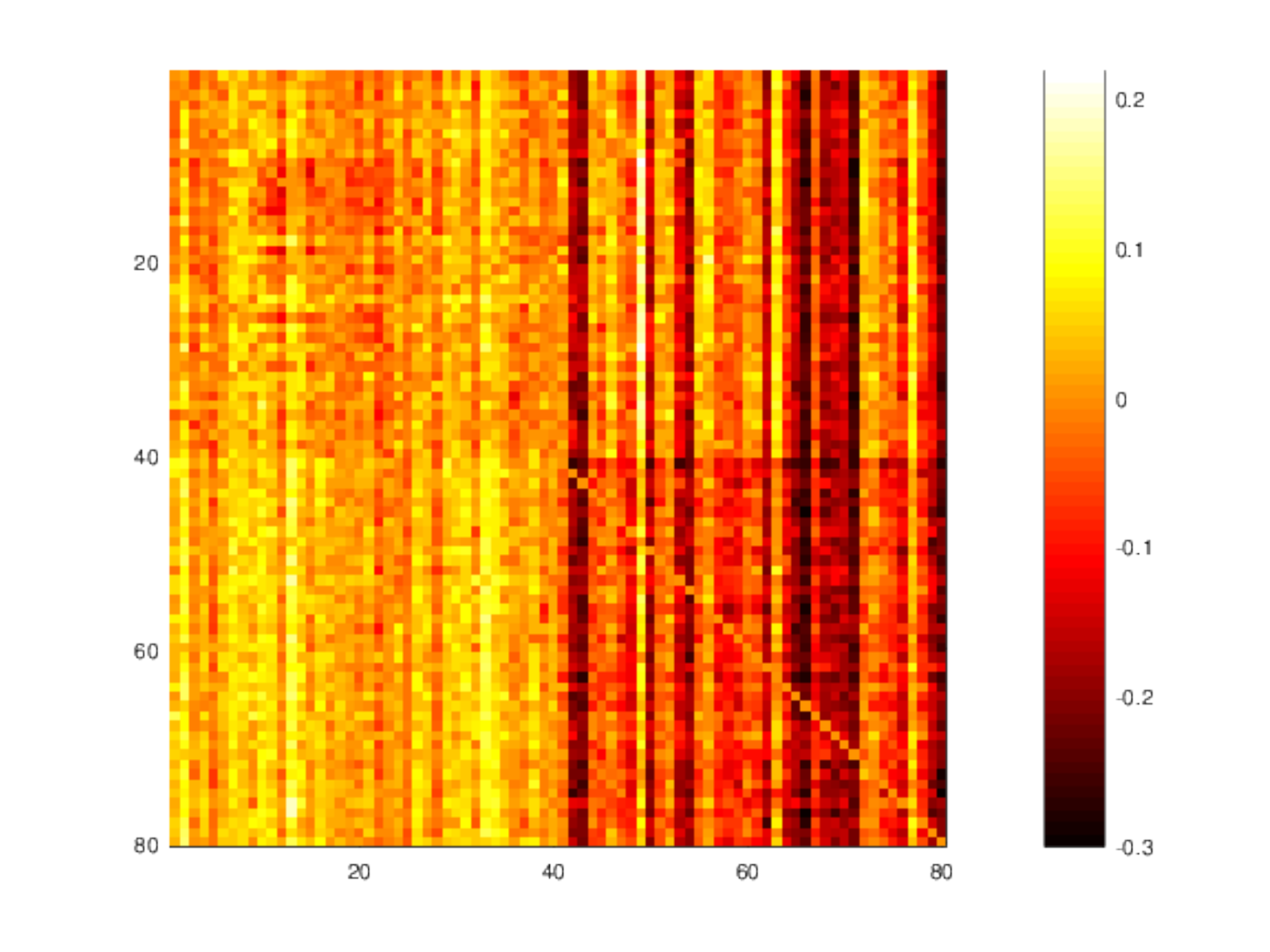}}
\caption{\footnotesize ETE matrix subtracting $M=1000$ shuffled time series, and kernel density resolution of $h=0.36$. (a) $k=l=1$; (b) $k=l=2$; (c) $k=l=3$; (d) $k=l=4$.}	
\label{ETE_L}
\end{figure}

On the other hand, the expression of TE~(Eq.~\ref{Transfer}) it is likely to be biased due to several factors as finite sample effects and
non-stationarity of data. Also time series that have more entropy, what is associated with higher volatility in finance, naturally transfer more entropy to the others. To reduce this bias, we use the effective transfer entropy~(ETE)~\cite{Marschinski:2002}
\begin{equation}
 ETE_{Y\rightarrow X} = TE_{Y\rightarrow X}(k,l) - \frac{1}{M} \sum^{M}_iTE_{Y_{(i)}\rightarrow X}(k,l)
\end{equation}
where $Y_{(i)}$ has been randomly shuffled from the time series $Y$.
By computing this quantity over all possible combinations of polarity and return time series of each country of Table~\ref{t.1}, we obtain an ETE matrix of dimensions $80 \times 80$.
In Fig.~\ref{ETE_L} are shown as a heatmap the ETE matrix results for $k=l=1,2,3,4$, and resolution $h=0.36$ (given by Eq. (\ref{dumb})), where the matrix elements have the same order as the partitioned correlation matrices~(see Fig.~\ref{super}). We have subtracted $M=1000$ random permutations of the corresponding $Y$ time series.
These results shows that most of the information flows from returns and polarity to returns exclusively, being the case $k=3$ where this phenomenon is clearer to the eye. \\

We used graph theory to transform the ETE matrix into a directed network.
Now each time series is represented by a node and the magnitude of the flow of information from one node to another by directed edge.
An important quantity of an undirected network is the number of edges connected to the node; or \emph{node degree}. For directed networks there are two related measures \emph{out node degree} $ND_{out}$ and \emph{in node degree} $ND_{in}$, which count the number of edges leaving and entering a node, respectively. 
Since we are interested in exploring the situation where the information flows exclusively to returns because it might open up new trading strategies, we will define the \emph{relative out node degree} as the ratio of the \emph{out node degree} between polarity and return nodes, that is $ND_{out}(polarity)/ND_{out}(returns)$.\\

\begin{figure}
\centering
\subfloat[]{\includegraphics[width=0.9\linewidth]{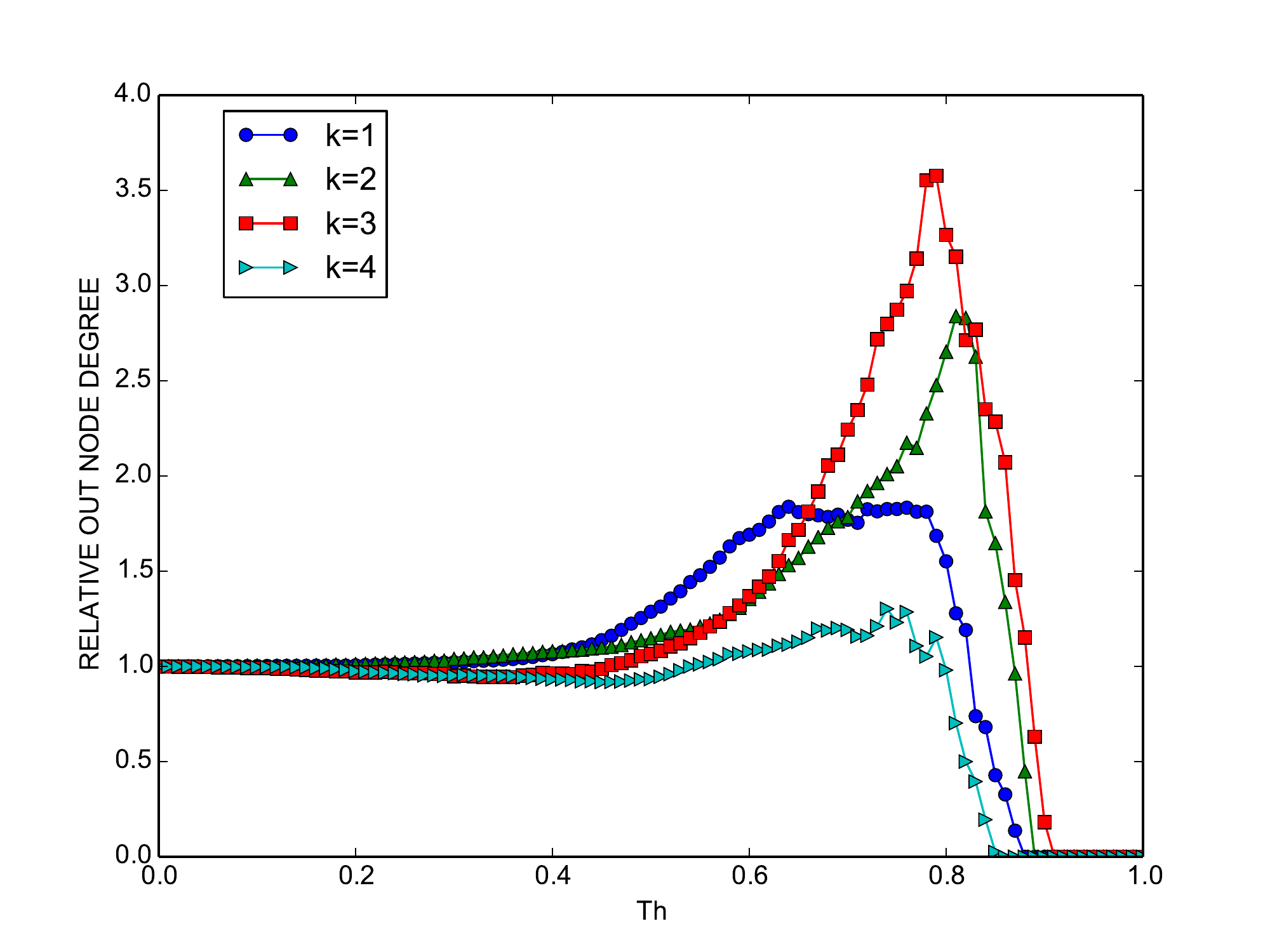}}\\
\subfloat[]{\includegraphics[width=0.9\linewidth]{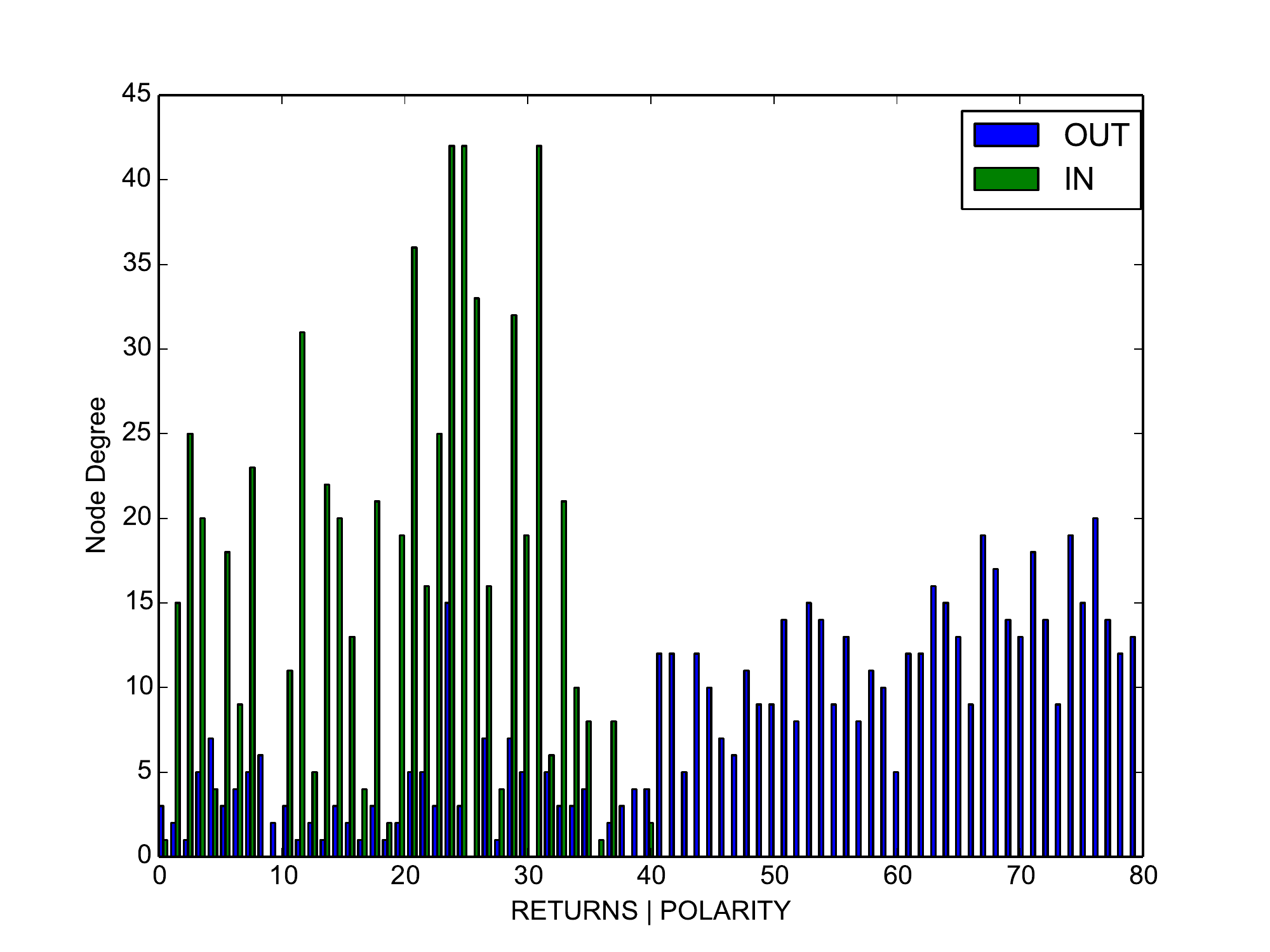}}
\caption{\footnotesize (a) \emph{relative out node degree} as a function of the normalized threshold $Th$. (b) \emph{in} and \emph{out node degree} for $k=3$,$Th=0.79$.}	
\label{threshold}
\end{figure}

In Fig.~(\ref{threshold})(a) we observe the \emph{relative out node degree} as a function of the rescaled range of values of ETE, which runs over the threshold $Th \in [0,1]$ for each $k=1,2,3,4$. We found the maximum at $Th=0.79$ for $k=3$; it is at this value when the nodes have more edges leaving from polarity to return nodes, and therefore it is a good indicator for analyzing the corresponding network. In Fig.~(\ref{threshold})(b) we plot bar graphs of the \emph{out} and \emph{in node degree} values for the $80$ nodes of our network at $k=3,Th=0.79$, where the first 40 nodes correspond to returns, and the last 40 to polarity nodes. 
It is very interesting that for these values of $k$ and $Th$, the polarity nodes are the only ones that send information to the whole network, and also  that the \emph{in node degree} for returns is bigger than for the polarity one.

\section{Conclusion}

We have found eigenvalues beyond the universal results of Wishart matrices for polarity and return data, which implies the presence of a global factor leading the set of indices and news as a unity.
Interestingly, we found that the largest eigenvalue of news and prices also share the same dynamics, result that sides with the one of behavioral finance.
The temporal analysis of IPR confirms the fact that each financial index participates significantly in the eigenvector associated to the largest eigenvalue. Notably, the data from NYT shows the same behavior. 

The results from the partitioned matrices in CWOE analysis, supports the fact of the existence of true correlations for world indices and news, showing that the correlation structure is preserved when adding with noise to the empirical data. Nevertheless, was not possible to characterize the cross-correlation between them because the small dimensions of the empirical correlation matrices, being necessary the use of the information theory approach. 
In this field, transfer entropy analysis revel us that for memory k = 3 and normalized threshold Th = 0.79, all the information flows to  return nodes. The last is the most practical result for trading purpose, suggesting a possible selection rule for an optimal historical news set, and showing new precise evidence in favor of Behavioral finance as a reliable economic paradigm. 

\section*{Acknowledgments}

We thank Vinayak and Fabio Ayres for fruitful discussions and advice as well as  Mois\'es Mart\'inez, Carlos Pacheco and Carlos Liz\'arraga for helpful comments. This work was partially supported by \emph{Consejo Nacional de Ciencia y Tecnolog\'ia}~(CONACyT) of Mexico, and Insper Institute of Education and Research at S\~ao Paulo, Brazil.

\bibliographystyle{ieeetr} 
\bibliography{main}

\end{document}